# Impact of the chemical structure on the dynamics of mass transfer of water in conjugated microporous polymers: A neutron spectroscopy study

Anne A. Y. Guilbert[1]*, Yang Bai[2], Catherine M. Aitchison[2], Reiner Sebastian Sprick[2], Mohamed Zbiri[3]*

[1]Department of Physics and Centre for Plastic Electronics, Imperial College London, Prince Consort Road, London SW7 2AZ, U.K.

[2]Department of Chemistry and Materials Innovation Factory, University of Liverpool, Crown Street, Liverpool L69 7ZD, U.K.

[3]Institut Laue-Langevin, 71 Avenue des Martyrs, Grenoble Cedex 9 38042, France

**ABSTRACT:** Hydrogen fuel can contribute as a masterpiece in conceiving a robust carbon-free economic puzzle if cleaner methods to produce hydrogen become technically efficient and economically viable. Organic photocatalytic materials such as conjugated microporous materials (CMPs) are potential attractive candidates for water splitting as their energy levels and optical bandgap as well as porosity are tunable through chemical synthesis. The performances of CMPs depend also on the mass transfer of reactants, intermediates and products. Here, we study the mass transfer of water ($H_2O$ and $D_2O$), and of triethylamine used as a hole scavenger for hydrogen evolution, by means of neutron spectroscopy. We find that the stiffness of the nodes of the CMPs is correlated with an increase in trapped water, reflected by motions too slow to be quantified by quasi-elastic neutron scattering (QENS). Our study highlights that the addition of the polar sulfone group results in additional interactions between water and the CMP, as evidenced by inelastic neutron scattering (INS), leading to changes in the translational diffusion of water, as determined from the QENS measurements. No changes in triethylamine motions could be observed within CMPs from the present investigations.

INTRODUCTION

The need for a renewable energy carrier has resulted in intense research over the last decades on the generation of hydrogen from water via water-splitting. Solar energy can be utilized to facilitate the water-splitting process using a photocatalyst. Most of the photocatalysts studied are inorganic,[1,2] but, since the first report on carbon nitrides as potential photocatalyst in 2009,[3] organic polymer photocatalysts have also been studied intensively.[4–6] Initially, carbon nitrides[3,7] were the main-focus but in recent years conjugated microporous polymer networks (CMPs),[8–10] linear conjugated polymers,[11–17] triazine-based frameworks,[18–21] covalent organic frameworks (COFs),[22–24] and molecular compounds[25,26] have also been proposed for sacrificial proton reduction half reaction. Activities that rival those obtained with inorganic systems have been achieved in some cases.[27-29] The interest in organic photocatalysts arises from the ease of synthesis of polymer photocatalysts via low-temperature routes that allow for precise control over the polymer sequence, hence, allowing for tailoring of their functionalities.[5,30]

Over the years, these studies have led to an understanding of the importance of several factors that result in high activity in polymer photocatalysts, such as light absorption,[8,31,32] driving-force for proton reduction and scavenger oxidation,[31] exciton separation[16,33] and, crystallinity.[34–36] Due to the hydrophobic nature of most polymeric photocatalyst surface, wetting seems to be particularly important.[37–39]. Several studies have shown that the introduction of polar groups results in materials with higher photocatalytic activities.[12,39–41] Large surface area to maximize the exposed surface to water can also be beneficial. Therefore, porous photocatalysts with high Brunauer–Emmett–Teller surface areas ($SA_{BET}$), namely COFs and CMPs,[42–44] have been developed for photocatalysis.[10,34,45–47] In a previous paper, we studied CMPs and their linear polymer analogues and we found that the porous materials do not always outperform their non-porous analogues.[45]

For porous materials, the interaction between the surface of the photocatalyst and water, which can be tuned by modifying the polarity of the photocatalyst,[34,45] as well as the size of the pores will impact the dynamics of water on the surface and within the material. If the water dynamics is particularly slow in comparison with the kinetics of the photocatalytic reaction, the increased surface area will benefit the overall activity little. However, very few studies have explored transport of water through organic materials and the interaction of water with the surface of these materials.[12,41] At the macroscopic scale, contact angle measurements with water and water sorption measurements give information about the wetting of particles,[12,41] and their available surface.[34,45] No kinetic information can be obtained by these techniques and specific interactions can only be inferred.

Neutron spectroscopy is a master technique of probe to study the guest-host dynamics, at the microscopic level. Quasi-elastic neutron scattering (QENS) and inelastic neutron scattering (INS) were applied recently to map in details the microstructural dynamics up to the nanosecond of the conjugated polymer poly(3-hexylthiophene), under both its regioregular and regiorandom forms.[48] QENS has found application in the study of the transport of lithium ions in inorganic electrodes for batteries,[49] gases in metal organic frameworks,[50] and the rotational dynamics of hydrogen adsorbed in covalent organic frameworks.[51] It has also been used to study water on the surface of oligonucleotide crystals and,[52] cages crystals.[53] We demonstrated previously that QENS can be used to study the water dynamics in CMPs.[45]

Here, we go a step further and combine QENS and inelastic neutron scattering (INS) to quantify the water dynamics in three CMPs and study the interactions between water and CMPs at the molecular level. We also report on the dynamics of the hole scavenger triethylamine (TEA) used for the sacrificial proton reduction half reaction. We select, as model systems, the previously reported F-CMP3, S-CMP3 and S-CMP1 (Figure 1 a,b).[45] The labelling of the systems is the same adopted in Reference .45, which also reported on the full relevant characterization to their fabrication and photocatalysis. Comparing F-CMP3 and S-CMP3 allows us to study the impact of introducing a polar sulfone group in the strut of the CMPs and comparing S-CMP3 with S-CMP1 enables us to study the impact of the network structure on the reactant (water and TEA) dynamics.

EXPERIMENTAL SECTION

The neutron scattering measurements were performed using the direct geometry, cold neutron, time-of-flight, time-focusing spectrometer IN6, and the hot-neutron, inverted geometry spectrometer IN1-Lagrange, at the Institut Laue-Langevin (ILL, Grenoble, France). About 300 mg of CMP samples were loaded into thin aluminium-made hollow cylindrical containers dedicated for neutron spectroscopy. An optimized sample thickness of 0.2 mm was considered, relevant to the minimization of effects like multiple scattering and absorption. The water was introduced in the container just before measurement and the container was tightly sealed, and an Indium wire was used as a gasket. The mass (the smallest mass of water equals about 50 mg) was recorded before and after each measurement. No evaporation was recorded. The water was fully evaporated between the measurements on the different instruments, and the same procedure was followed for new measurements. All measurements are performed under vacuum. The QENS spectra were collected on IN6 using an ILL orange cryostat at 2, 200 and 300 K, and an incident neutron wavelength of 5.12 Å ($E_i \approx 3.12$ meV), offering an optimal energy resolution at the elastic line of ~ 0.07 meV. Standard corrections including detector efficiency calibration and background subtraction were performed. A vanadium sample was used to calibrate the detectors and to measure the instrumental resolution under the same operating conditions. At the used wavelength ($\lambda_i$= 5.12 Å), the IN6 angular detector coverage (~ 10 - 114°) corresponds to a Q-range of ~ 0.2–2.1 Å$^{-1}$. The data reduction and analysis were done using ILL software tools. For the QENS spectra, different data sets were extracted either by performing a full Q-average in the (Q, E) space to get the scattering function S(E, T) or by considering Q-slices to study the S(Q, E, T). The INS spectra, in terms of the generalized density of states (GDOS),[54] were collected using both IN6 and IN1-Lagrange. On IN6, this was done concomitantly with the acquisition of the QENS data, in the up-scattering, neutron energy-gain mode, and the one-phonon GDOS were extracted, within the incoherent approximation framework.[55-57] On IN1-Lagrange, the GDOS spectra were collected in the down-scattering, neutron energy-less mode at 10 K, using a closed cycle refrigerator, with the fixed final analyzer energy of 4.5 meV. The incident energy was varied in a stepwise manner via Bragg scattering from a copper monochromator crystal. In this work, using the doubly focused Cu(220) monochromator setting, the incident energy was ~210–3500 cm$^{-1}$, leading after subtraction of the fixed final energy value (4.5 meV) to an accessible energy transfer range of ~180–3500 cm$^{-1}$, hence covering the full molecular vibrational frequencies.

RESULTS AND DISCUSSION

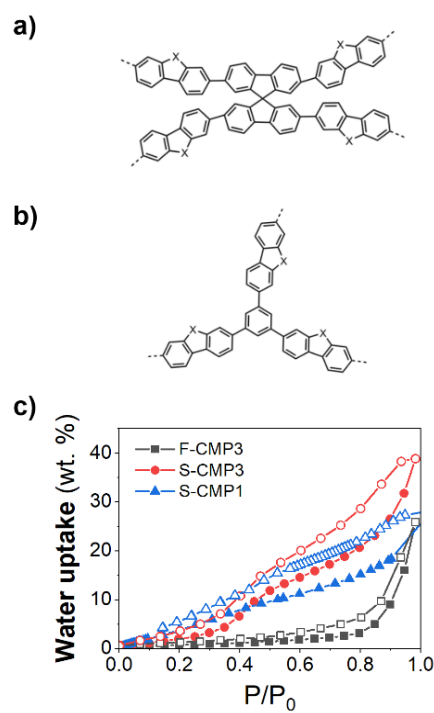

Figure 1. Schematic illustration of the chemical structure of (a) F-CMP3 (X=CH$_2$), S-CMP3 (X=SO$_2$), and (b) S-CMP1

(X=SO$_2$). (c) Water uptake measurements as a function of the relative pressure P/P$_0$ (P$_0$ is the saturation pressure of water) at 20.0 °C of F-CMP3, S-CMP3 and S-CMP1, showing the evolution of both adsorption (solid symbols) and desorption (open symbols) processes.

The CMPs were synthesized using previously reported methods.[45] All materials were found to be porous to nitrogen with Brunauer-Emmett-Teller surface areas ($SA_{BET}$) determined to be 596 m$^2$ g$^{-1}$ for F-CMP3, 431 m$^2$ g$^{-1}$ for S-CMP3 and 508 m$^2$ g$^{-1}$ for S-CMP1. The relatively high $SA_{BET}$ for all three CMPs may allow for water penetration into the network as water sorption measurements show water uptake for all the CMPs (Figure 1 c); however, condensation on the surface cannot be rules out. S-CMP1 adsorbs water at lower relative pressure than both S-CMP3 and F-CMP3 but S-CMP3 uptake is higher overall. F-CMP3 adsorbs only at very high relative pressure.

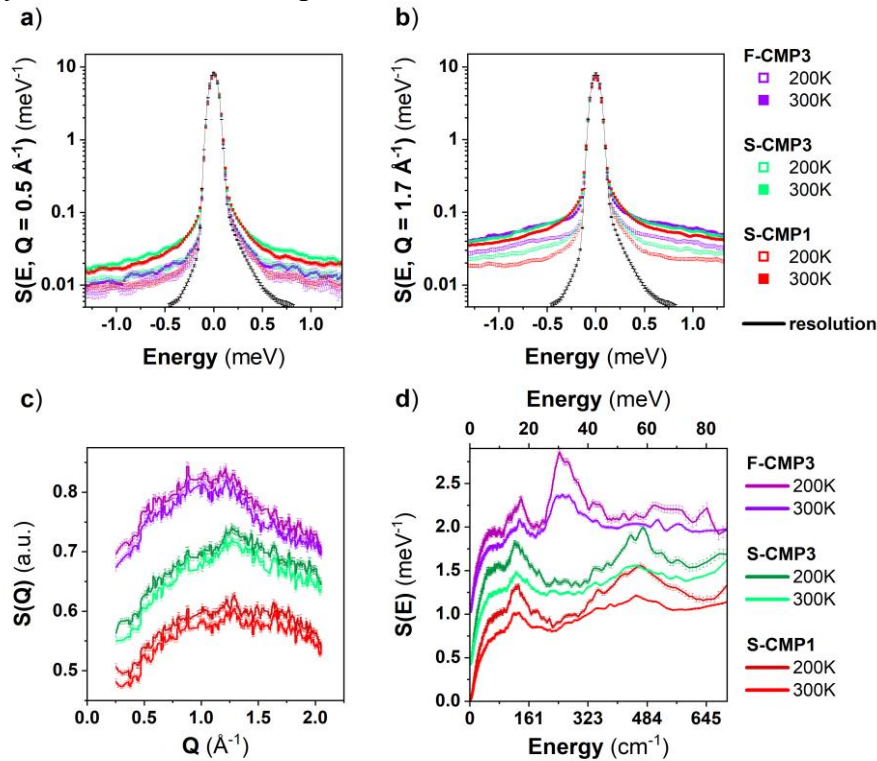

Figure 2. Top panel: Q-dependent QENS spectra of dried F-CMP3, S-CMP3 and S-CMP1 at 200K and 300K: a) Q = 0.5 Å$^{-1}$ and b) Q = 1.7 Å$^{-1}$. The instrumental resolution function of IN6 is measured by quenching S-CMP3 at 2K and is represented by the narrow black solid elastic line. Bottom panel: c) Neutron diffractograms of F-CMP3, S-CMP3 and S-CMP1, extracted from the same IN6 measurements at 200 K and 300 K. d) Generalized phonon density of states (GDOS) of F-CMP3, S-CMP3 and S-CMP1 at 200 and 300K, also obtained from the same IN6 measurements.

Figure 2 (a-c) shows the temperature evolution of the Q-dependence of the QENS spectra of F-CMP3, S-CMP3 and S-CMP1. The QENS spectra of the three CMPs, on the accessible instrumental energy window, is mainly elastic with a background increasing with both temperature and Q, and a small quasi-elastic contribution with a half-width-half-maximum (HWHM) of about 0.2 meV. The overall Q-dependence of F-CMP3 is more pronounced than for both S-CMP1 and S-CMP3. At 200K, only F-CMP3 presents a Q-dependence of the quasi-elastic contribution. At 300K, the quasi-elastic contribution is broader for S-CMP1 and S-CMP3 as can be clearly seen in Figure 2 b. Indeed, the background for all the three CMPs is similar at 300K and at Q = 1.1 Å$^{-1}$ but the quasi-elastic contribution is narrower in the case of F-CMP3. The Q-dependence of the background of F-CMP3 is stronger at 300K than for S-CMP3 and S-CMP1, which is consistent with what is observed at 200K, indicating that the degree of freedom captured at 200K by the instrumental energy window becomes too fast at 300K to be properly resolved and contribute to the background. Hydrogen has a large incoherent neutron cross section in comparison with oxygen and therefore, the Q-dependent motion seen at 200K and linked with the background at 300K is assigned to a motion related to the -C(CH$_3$)$_2$ group of F-CMP3. Within the energy window of the instrument, the rotational motion of the entire linker is likely to be captured at 300K. The presence of the -C(CH$_3$)$_2$ group seems to induce a frustration of this motion in comparison with the sulfone group -SO$_2$. This could be explained by the larger -C(CH$_3$)$_2$ group more likely to create a steric hindrance. Interestingly, no strong differences in the QENS spectra between S-CMP1 and S-CMP3 related to the difference of nodes are observed at those temperatures, within this instrumental energy window. The CMPs are reported to be largely amorphous as measured by powder X-Ray diffraction.[45] All the CMPs feature a broad Bragg peak around 1.3 Å$^{-1}$ as measured by X-Ray diffraction[45] and as observed by neutron diffraction in Figure 2 e .Further neutron diffractograms down to 2K are presented in Figure S9 in the Supporting Information. No differences as a function of temperature is observed for all the CMPs. The generalized density of states (GDOS)[1] spectra of S-CMP1 and S-CMP3 exhibit similar vibrational aspects, both in intensity and profile (Figure 2 f), and noticeable differences compared to F-CMP3. Indeed, the vibrational band observed around 480 cm$^{-1}$ in S-CMP1 and S-CMP3 is absent in F-CMP3, while the band around 250 cm$^{-1}$ in F-CMP3 is absent in S-CMP1 and S-CMP3. Thus, the features around 250 and 480 cm$^{-1}$ can be assigned to vibrational modes involving the -CH$_2$ group and the sulfone group -SO$_2$, respectively. Upon cooling from 300 to 200 K, the peaks at 250 and 480 cm$^{-1}$ in F-CMP3 and S-CMP3, respectively, exhibit a pronounced narrowing while the narrowing of the peak at 480 cm$^{-1}$ for S-CMP1 is less pronounced than in S-CMP3 (Figure 2 e). This can be explained by the different nodes in F-CMP3 and S-CMP3 in comparison with S-CMP1. The spiro node is expected to be more rigid leading to a more ordered structure in F-CMP3 and S-CMP3 than in S-CMP1. Although the overall GDOS is more impacted by the presence of the sulfone group than the differences in nodes, INS proves to be very sensitive to the subtle differences of the chemical structures of F-CMP3, S-CMP3 and S-CMP1. The energy range up to 700 cm$^{-1}$, from the cold-neutron measurements using IN6, likely covers the external (phonon) modes (Figure 2 f).

In order to probe a full spectrum including the internal (molecular) degrees-of-freedom, we went a step further and performed measurements on the hot-neutron IN1-Lagrange spectrometer on F-CMP3 and S-CMP3 (Figure 3). We focus on comparing the vibrational response of F-CMP3 and S-CMP3 to further understand the impact of the sulfone group on to the molecular vibrations in terms of interaction with water. To get a resolved and structured molecular vibrational spectrum, the spectra were collected at 10 K in order to considerably reduce the temperature-induced Debye-Waller effect. This allowed to extend the accessible energy range of the IN6 spectra to higher energies on IN1-Lagrange, hence leading to cover the full molecular vibrational range, up to the C-H stretch band around 3600 cm$^{-1}$. The bands at 250 and 480 cm$^{-1}$ are also well captured in the IN1-Lagrange measurements. Compared to IN6, where measurements were performed at 200 and 300 K, decreasing the temperature to 10 K on IN1-Lagrange enabled us to better resolve both the features at 250 and 480 cm$^{-1}$.

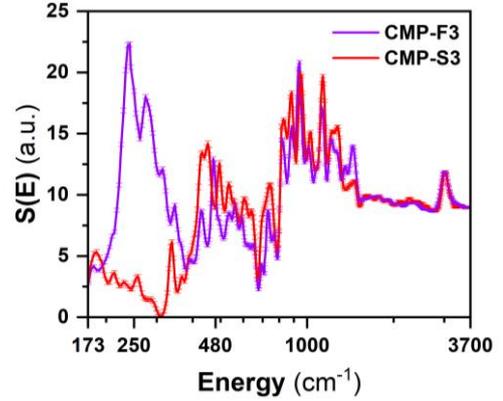

Figure 3. GDOS of F-CMP3 and S-CMP3, at 10 K, measured on the hot-neutron spectrometer IN1-Lagrange, allowing to probe the full molecular vibrational spectrum up to 3700 cm$^{-1}$ (~ 459 meV).

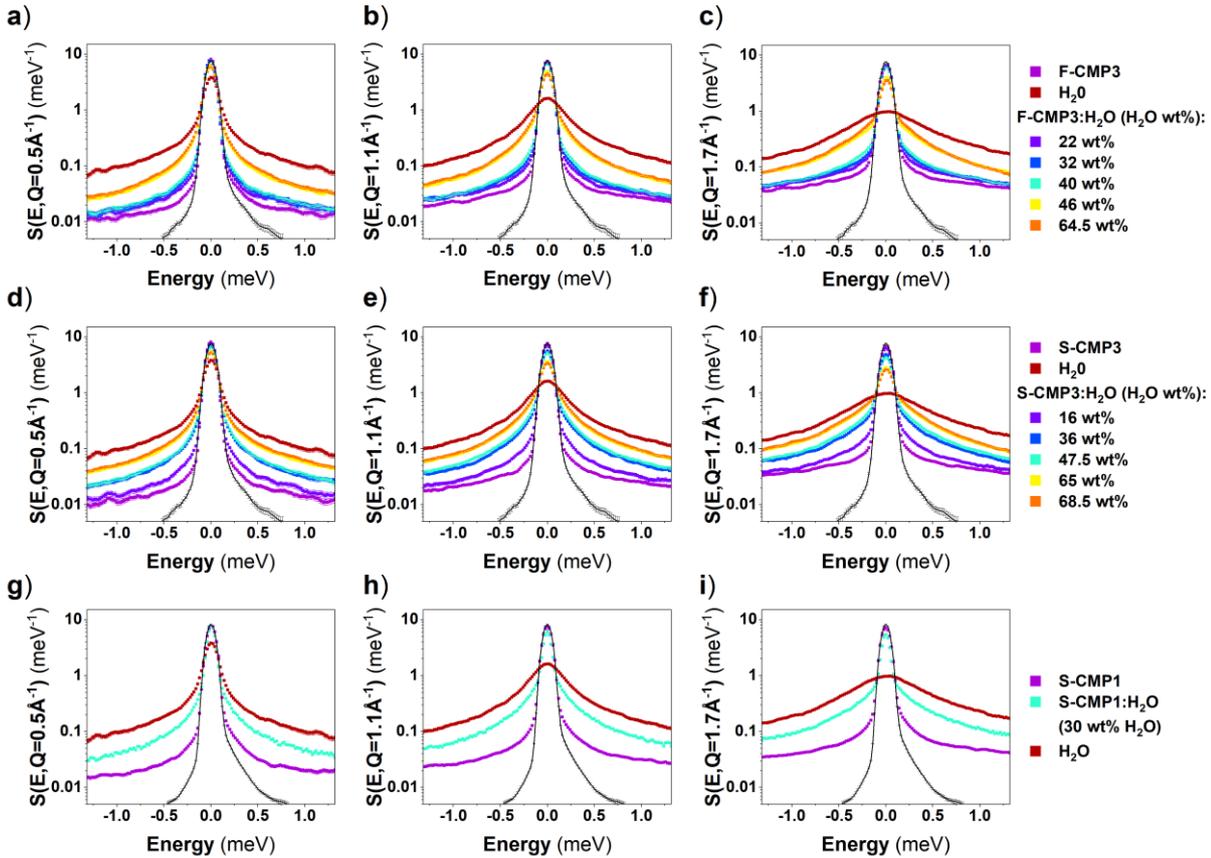

Figure 4. Q-dependent room-temperature QENS spectra of dried CMPs, bulk H$_2$O and CMPs mixed with different amounts of H$_2$O, CMPs:H$_2$O, for: (a-c) F-CMP3, (d-f) S-CMP3 and (g-i) S-CMP1, at (a,d,g) Q = 0.5 Å$^{-1}$, (b,e,h) Q = 1.1 Å$^{-1}$, and (c,f,i) Q = 1.7 Å$^{-1}$. The instrumental resolution function is measured by quenching S-CMP3 at 2K, and is represented by the narrow black solid elastic line.

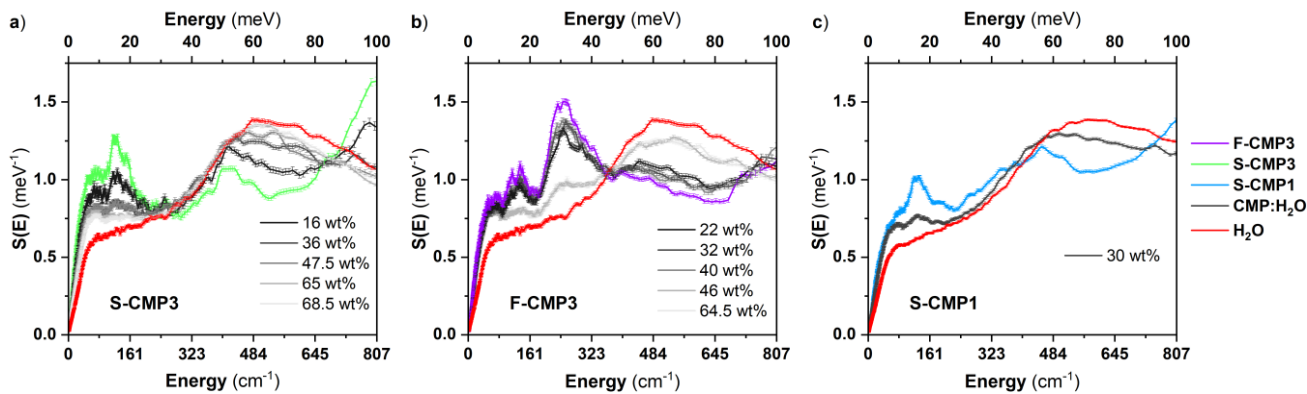

Figure 5. GDOS from IN6 measurements at 300K of a) F-CMP3, b) S-CMP3 and c) S-CMP1 mixed with different amount of $H_2O$.

Figure 4 compares the Q-dependence of the QENS spectra of the three CMPs mixed with $H_2O$ (CMPs:H2O) with the Q-dependence of the QENS spectra of dried CMPs and bulk $H_2O$. At higher $H_2O$ concentrations, the spectra are expected to be dominated by the signal of $H_2O$, as the neutron incoherent cross section of $H_2O$ is larger than the neutron incoherent cross section of the CMPs (Table S1 in Supporting Information). The presence of the sulfone group $-SO_2$ in S-CMP3 lowers the neutron incoherent cross section with respect to the neutron incoherent cross section of F-CMP3, thus, the $H_2O$ contribution to the overall QENS spectra is dominating the QENS signals of S-CMP3:$H_2O$ and F-CMP3:$H_2O$ for concentrations above 16 wt% and 40 wt%, respectively. As mentioned above, the QENS spectra of CMPs are mainly elastically shaped, and an increased elastic contribution is observed, in comparison with bulk $H_2O$, even for the highest $H_2O$ concentrations. Water can either bound to the CMP, be strongly adsorbed on the surface of the pores leading to strong hindrance of water motions and diffusion or can be free to diffuse. We will refer to these three types of water as bound water, constrained water and free water, respectively. Trapped water will be used as a loose term encompassing both bound and constrained water. This increased elastic contribution could originate from the CMPs signals or may be due to the presence of bound water. For simplicity, we use hereafter the wording bound water to refer to both water bound to CMPs and water with motions too slow to be captured by the instrument. The spectra of water in CMPs appear to be narrower than the bulk water signal, pointing towards the presence of constrained or trapped water. The CMPs spectra may change with the presence of water, and may dominate the changes in QENS spectra at lower $H_2O$ concentrations. It cannot be ruled out without a further analysis that the observed changes in the QENS spectra of CMPs:$H_2O$ with respect to the dried CMPs are a combination of a change in the QENS signals of both the CMPs and water. To gain insights into the specific behavior of water in the different CMPs, we further exploit the GDOS of water and water-mixed CMPs, from IN6 INS measurements. Figure 5 shows the GDOS of the dried CMPs, of the CMPs mixed with $H_2O$ and of bulk $H_2O$. The broad peak around 80 meV of bulk $H_2O$ is assigned to the libration of water.[58,59] It can be fitted by a combination of 3 gaussians representing the rock, wag, and twist modes of water as presented in Figure 6 for bulk water.

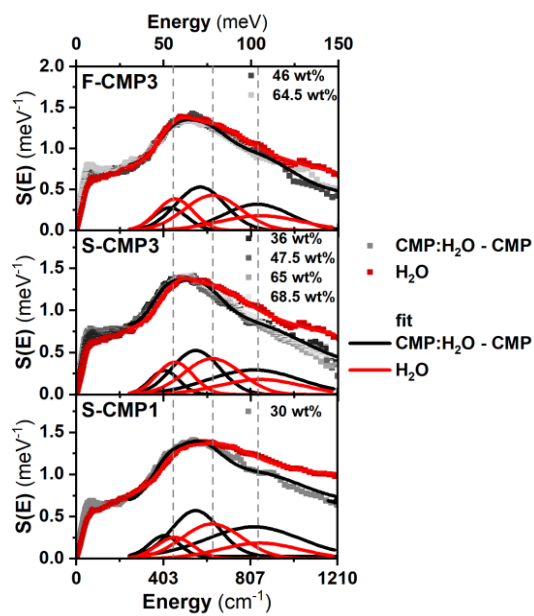

Figure 6. GDOS of bulk reference $H_2O$ and $H_2O$ in (top) F-CMP3, (middle) S-CMP3 and (bottom) S-CMP1. The GDOS of $H_2O$ in the CMP samples is presented here as the difference of the water-mixed CMPs (either F-CMP3:$H_2O$, S-CMP3:$H_2O$ or S-CMP1:$H_2O$) and dried CMPs (either F-CMP3, S-CMP3 or S-CMP1, respectively). The broad peak at around 80 meV is assigned to the libration band of water and is fitted for bulk water (red line) and for the difference of the water-mixed CMPs and dried CMPs (blacklines) with a combination of 3 gaussians representing the rock, wag, and twist modes of water.[2,3]

In Figure 6, bulk $H_2O$ is presented as a "reference", compared to the difference of the GDOS of the water-mixed CMPs and dried CMPs. The intensity of the low-energy feature of water, at ∼ 7 meV, increases for all CMPs although more significantly for F-CMP3. This could reflect a change in organization of water, especially in the hydration monolayer of all the CMPs. The difference in GDOS of the water-mixed CMPs and dried CMPs is fitted similarly to the bulk water with 3 gaussians. The comparison gaussian-wise between bulk water (red line) and the difference in GDOS (black lines) clearly highlights some hindrance and change in the vibrational distribution of the librational degrees-of-freedom of $H_2O$ in the CMPs. In order to quantify this hindrance for the respective CMPs, we calculate the weighted librational peak position (WLPP)[59] as follow:

$$WLPP = \frac{A_{G1} \times x_{G1} + A_{G2} \times x_{G2} + A_{G3} \times x_{G3}}{A_{G1} + A_{G2} + A_{G3}} \quad (1)$$

where $A_{Gn}$ and $x_{Gn}$ are the area and the center of the gaussians, respectively. A higher WLPP value indicates a higher energy required to excite the water libration mode or in other words, a lower water librational mobility. The WLLP is higher for S-CMP3 and S-CMP1 in comparison with F-CMP3 although a smaller hindrance of the librational water mobility is still observed for F-CMP3 (Figure 7).

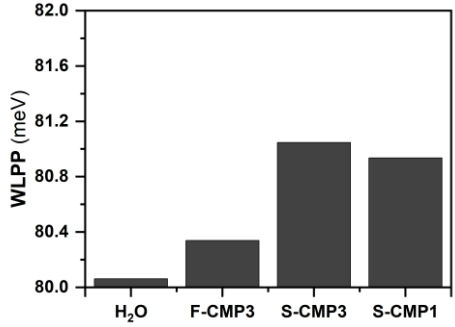

Figure 7. Weighted librational peak position (WLPP) in meV for $H_2O$, F-CMP3, S-CMP3 and S-CMP1.

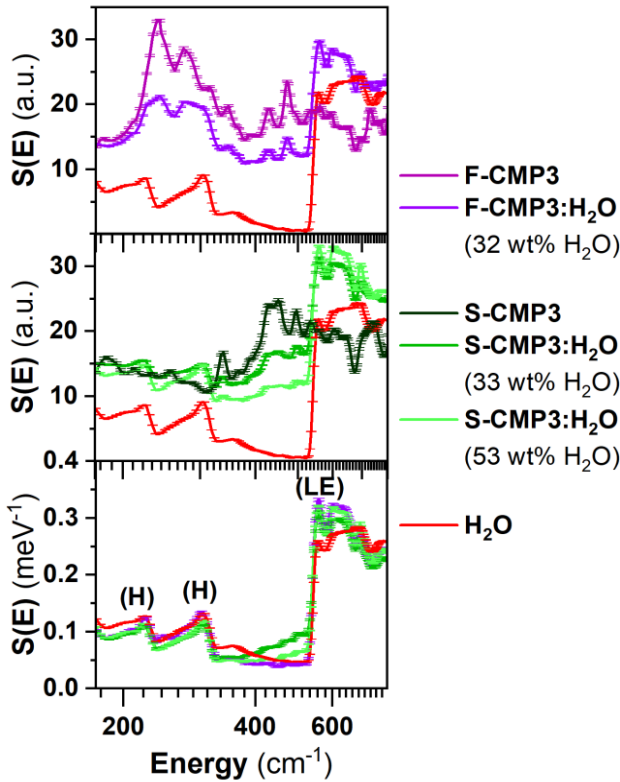

Figure 8. GDOS spectra, from INS measurements at 10K using IN1-Lagrange, of (top) F-CMP3 and (middle) S-CMP3 with $H_2O$. (bottom) The GDOS of bulk reference $H_2O$ and $H_2O$ in the F-CMP3 and S-CMP3, taken as the difference of the water-mixed CMPs (either F-CMP3:$H_2O$ or S-CMP3:$H_2O$) and dried CMPs (either F-CMP3 or S-CMP3, respectively). The peaks labelled as (H) are assigned to the hydrogen-bond bending and stretching[60] and as (LE) to refer to the librational edge.[52,61]

To better resolve the librational motion of water, we further perform measurements for S-CMP3 and F-CMP3 with $H_2O$ at 10K (Figure 8) to reduce significantly the Debye-Waller effect, using IN1-Lagrange. The vibrational spectra of the water-mixed CMP samples include contribution of both the GDOS of the dried CMPs and reference bulk $H_2O$ (ice down to 10K). This points towards the presence of free water at the probed concentration. By subtracting the contribution of the dried F-CMP3 and S-CMP3 from the two water-mixed CMP form (Figure 8-bottom), it appears that the GDOS of $H_2O$ in both F-CMP3 and S-CMP3 deviate from the GDOS of the reference bulk $H_2O$. The features at 225 cm$^{-1}$ and 300 cm$^{-1}$ are assigned to the hydrogen bond bending and stretching components of bulk ice while the edge at 300 cm$^{-1}$ is assigned to the libration edge of ice. No differences are observed in the region 200-320 cm$^{-1}$. A broad feature around 500 cm$^{-1}$ is observed for $H_2O$ in S-CMP3 and its magnitude varies with water concentration. The sharp libration edge of ice is observed for $H_2O$ in both S-CMP3 and F-CMP3; however, an extra contribution for both materials is seen at 600 cm$^{-1}$. From this vibrational study, it can be concluded that the broad features at 500 cm$^{-1}$, in the case of S-CMP3, and the additional feature at 600 cm$^{-1}$ for both S-CMP3 and F-CMP3, are associated with interfacial water. The absence of extra features in the region dominated by the stretching and bending of weak hydrogen bonding reveals that the structures of interfacial water are perturbed considerably from the bulk state for both CMPs. Furthermore, the extra feature at about 500 cm$^{-1}$ for S-CMP3 appears at a frequency where, in the dried CMPs, a more pronounced band is observed for S-CMP3 compared to F-CMP3 (3), thus, indicating a specific interaction between the sulfone group and water. To summarize, the hindrance of the librational degrees-of-freedom of water in S-CMP3 and S-CMP1 is a clear indication of the transition from free water to constrained/trapped water and/or bound water. The changes observed at 10K for both F-CMP3 and S-CMP3 reflect the presence of bound water, as well as an additional interaction between the sulfone group and water.

The behavior of water can further be explored and quantified, by fitting the QENS data. The dynamical structure factor of water $S_{water}(Q, \omega)$ is expressed as a convolution of the dynamical structure factors of the vibrational $S_V(Q, \omega)$, translational $S_T(Q, \omega)$ and rotational motions of water $S_R(Q, \omega)$.[62]

$$S_{water}(Q, \omega) = S_V(Q, \omega) \otimes S_T(Q, \omega) \otimes S_R(Q, \omega) \quad (2)$$

The $S_V(Q, \omega)$ component is mainly elastic with a background due to vibrations (inelastic contributions) and thus, can be written as $A(Q)\delta(\omega) + B(Q)$. $A(Q)$ is proportional to the Debye-Waller factor, $\delta(\omega)$ is a Dirac function and $B(Q)$ is the background due to vibrations. $S_T(Q, \omega)$ is represented by a single Lorentzian function $\mathcal{L}(\omega, \Gamma_T(Q))$ of HWHM $\Gamma_T(Q)$. We use the well-known Sears formalism[63,64] to describe $S_R(Q, \omega)$:

$$S_R(Q,\omega) = j_0^2(Qa)\delta(\omega) + 3j_1^2(Qa)\mathcal{L}\left(\omega, \frac{\hbar}{3\tau_R}\right) + 5j_2^2(Qa)\mathcal{L}\left(\omega, \frac{\hbar}{\tau_R}\right) \quad (3)$$

where $j_k$ is the k$^{th}$ Bessel function, $a$ is the radius of rotation, taken to be the O-H distance in the water molecule (0.98 Å), $\hbar$ is the reduced Planck constant and $\tau_R$ denotes the relaxation time of rotational diffusion. Up to a momentum transfer Q = 1.1 Å$^{-1}$, the third term can be neglected but at Q = 1.7 Å$^{-1}$, the first term becomes smaller than the third term (see Table S2 in Supporting Information). Thus, we keep the three terms, given the Q-range of the instrument, and $\tau_R$ is shared through each dataset during the fit to minimize the error on $\Gamma_T(Q)$. The QENS spectrum of water $I_{water}(Q,\omega)$, taking into account the resolution of the instrument $R(\omega)$, can be expressed as:

$$I_{water}(Q,\omega) = S_{water}(Q,\omega) \otimes R(\omega) = A(Q)\left\{\left(j_0^2(Qa)L(\omega,\Gamma_T(Q)) + 3j_1^2(Qa)L\left(\omega,\Gamma_T(Q) + \frac{\hbar}{3\tau_R}\right) + 5j_2^2(Qa)L\left(\omega,\Gamma_T(Q) + \frac{\hbar}{\tau_R}\right)\right) \otimes R(\omega)\right\} + B(Q) \quad (4)$$

The Q-dependence of $\Gamma_T$ is expected to follow the random-jump-diffusion model:[65]

$$\Gamma_T(Q) = \frac{D_T Q^2}{1 + D_T \tau_T Q^2} \quad (5)$$

where $D_T$ and $\tau_T$ are the translational diffusion constant and the residence time of the translational diffusion, respectively.

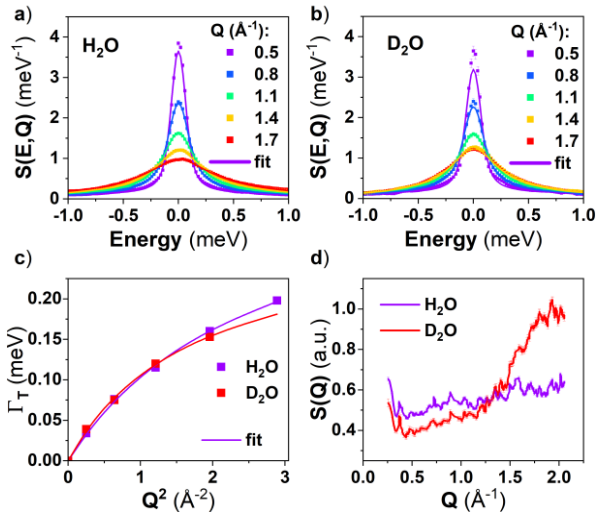

Figure 9. (a-b) Measured (scatter points), and the associated fit (solid line) using the described model, room temperature Q-dependent QENS spectra of a) H$_2$O and b) D$_2$O. c) HWHM of the Lorentzian representing the translational diffusion of water extracted from the fits of the QENS spectra as a function of Q$^2$ (scatter points) and fits using the random-jump-diffusion-model (solid line). d) Diffractograms of H$_2$O and D$_2$O extracted from the present IN6 measurement.

This model fits reasonably well both H$_2$O and D$_2$O data ($\chi^2$ = 0.46 and 0.45, respectively) – at the exception of Q = 1.7 A$^{-1}$ for D$_2$O (Figure 9 a-b), due to the pair distribution function of D$_2$O exhibiting a strong Bragg peak around Q = 1.7 A$^{-1}$ (Figure 9 d). We find a relaxation time $\tau_R$ = 0.940 ps for both H$_2$O and D$_2$O, a residence time $\tau_T$ = 1.736 ps for H$_2$O and 2.257 ps for D$_2$O, and a diffusion coefficient $D_T$ = 2.2 10$^{-5}$ cm$^2$.s$^{-1}$ for H$_2$O and 2.5 10$^{-5}$ cm$^2$.s$^{-1}$ for D$_2$O (Figure 9 c). This compares well with the literature where the residence time for the rotation and translation are both 1.1 ps and the diffusion coefficient is 2.3 10$^{-5}$ cm$^2$.s$^{-1}$.[7] The remaining fitting parameters can be found in Table S3 in Supporting Information.

To fit the QENS signals of the CMPs mixed with H$_2$O, the above model can further be formulated as:

$$I(Q,\omega) = (S_{CMP}(Q,\omega) + S_{H2O}(Q,\omega)) \otimes R(\omega) = C \times I_{CMP}(Q,\omega) + (1-C) \times A(Q) \times \left\{\left(j_0^2(Qa)L(\omega,\Gamma_T) + 3j_1^2(Qa)L\left(\omega,\Gamma_T(Q) + \frac{\hbar}{3\tau_R}\right) + 5j_2^2(Qa)L\left(\omega,\Gamma_T(Q) + \frac{\hbar}{\tau_R}\right)\right) \otimes R(\omega)\right\} + B(Q) \quad (6)$$

where $I_{CMP}(Q,\omega)$ is the measured normalised signal of the CMP and $C$ is the contribution of the CMP to the signal of CMP:H$_2$O, which can, in principle, be calculated from Table S1 in Supporting Information. Three possible types of water can be present: bound water, constrained water, and free water. The presence of bound water can lead to an extra elastic contribution. The QENS spectra of the dried CMPs being mainly elastic, the difference between $C$ extracted from the fit and calculated from Table S1 in Supporting Information is used to estimate the amount of bound water. $A(Q)$ is fixed here and the corresponding values are taken to be equal to those extracted from the fit of the free water (see Supporting Information Table S2). To avoid over parametrization, treatments of constrained water and free water are averaged. Thus, the diffusion coefficient extracted from Figure 10 represents an upper limit to the diffusion coefficient of constrained water.

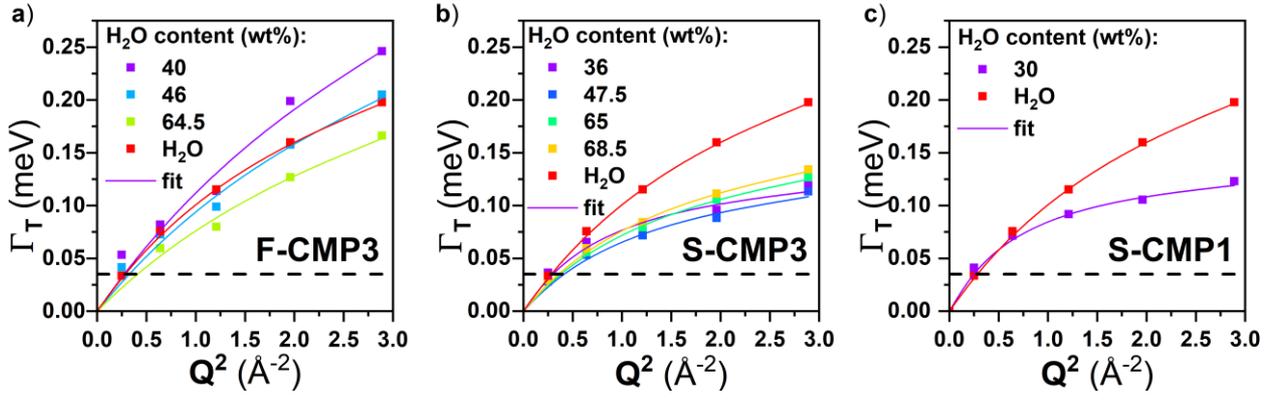

Figure 10. HWHM of the Lorentzian representing the translational diffusion of water extracted from the fits of the QENS spectra (scatter points) as a function of $Q^2$ and fits using the random-jump-diffusion-model (solid line) for a) F-CMP3:$H_2O$, b) S-CMP3:$H_2O$, both at different $H_2O$ concentrations and c) S-CMP1:$H_2O$. The horizontal dashed line represents the instrumental resolution.

Table 1. Main parameters obtained from the above described fitting procedure. The remaining fitting parameters can be found in Supporting Information (Table S4-S6). The amount of bound water is calculated from: the experimental water content, the value of $C$ from the fit and the estimated neutron incoherent cross sections. The expected $C$ value is inferred from the experimental water content and the estimated neutron incoherent cross sections with the assumption that in this case $C$ is solely linked with the contribution of the CMP to the overall QENS spectra. The translational residence time and diffusion coefficients are extracted from the fits of the HWHM, obtained from fitting the QENS data, as a function of $Q^2$ as shown in Figure 6.

|  | water content (wt%) | $C$ expected | $C$ from fit | bound water (wt%) | $\tau_R$ (ps) | $D_T$ ($10^{-5}$ cm$^2$ s$^{-1}$) | $\tau_T$ (ps) |
|---|---|---|---|---|---|---|---|
| $H_2O$ | 100.0 | 0.00 | 0.00 | 0.00 | 0.94 | 2.15 | 1.74 |
| F-CMP3 | 22.0 | 55.1 | 94.3 | 19.8 | N.A. | N.A. | N.A. |
|  | 32.0 | 42.4 | 90.8 | 28.3 | N.A. | N.A. | N.A. |
|  | 40.0 | 34.2 | 87.8 | 35.2 | 1.32 | 2.10 | 1.03 |
|  | 46.0 | 22.1 | 55.9 | 21.6 | 1.32 | 1.81 | 1.34 |
|  | 64.5 | 38.1 | 45.3 | 36.5 | 1.32 | 1.46 | 1.66 |
| S-CMP3 | 16.0 | 55.4 | 90.9 | 13.2 | N.A. | N.A. | N.A. |
|  | 36.0 | 29.6 | 65.3 | 22.2 | 1.32 | 2.60 | 4.49 |
|  | 47.5 | 20.7 | 56.4 | 28.9 | 1.30 | 1.76 | 4.11 |
|  | 65.0 | 11.3 | 29.5 | 27.6 | 1.13 | 1.79 | 3.34 |
|  | 68.5 | 9.8 | 26.5 | 28.7 | 1.14 | 1.89 | 3.14 |
| S-CMP1 | 30.0 | 34.0 | 52.8 | 10.9 | 0.97 | 3.21 | 4.45 |
| $D_2O$ | 100.0 | 0.00 | 0.00 | 0.00 | 0.94 | 2.52 | 2.26 |
| F-CMP3 | 32.0 | 96.8 | 96.7 | 0.0 | N.A. | N.A. | N.A. |
|  | 48.0 | 93.9 | 97.2 | 33.1 | N.A. | N.A. | N.A. |
|  | 65.0 | 88.4 | 94.6 | 49.8 | N.A. | N.A. | N.A. |
| S-CMP3 | 33.0 | 95.1 | 98.2 | 23.8 | N.A. | N.A. | N.A. |
|  | 48.0 | 91.2 | 93.9 | 21.7 | N.A. | N.A. | N.A. |
|  | 67.0 | 82.6 | 82.4 | 0.0 | N.A. | N.A. | N.A. |
| S-CMP1 | 30.0 | 95.5 | 82.7 | 0.0 | N.A. | N.A. | N.A. |

The model fits well with the higher $H_2O$ concentrations where the $H_2O$ signal dominates the QENS spectra (Supporting Information: Figures S1, S3 and S5). The water dynamics is impacted when mixed with S-CMP3 and S-CMP1, while it is unchanged in F-CMP3 (Figure 10). The residence time $\tau_T$ of the translational motion increases from about 1 ps for bulk water to about 4 ps when water is mixed with both S-CMP1 and S-CMP3. The diffusion coefficient value decreases with respect to bulk water when water is mixed with S-CMP3 but seems similar when mixed with S-CMP1 (Table 1). We quantify the amount of bound water for all CMPs from the difference between the expected $C$ calculated from Table S1 in Supporting Information, and $C$ extracted from the fit. F-CMP3 exhibits the largest amount of bound water with about 30 wt%, S-CMP3 has about 25 wt%, while S-CMP1 shows about 10 wt%.

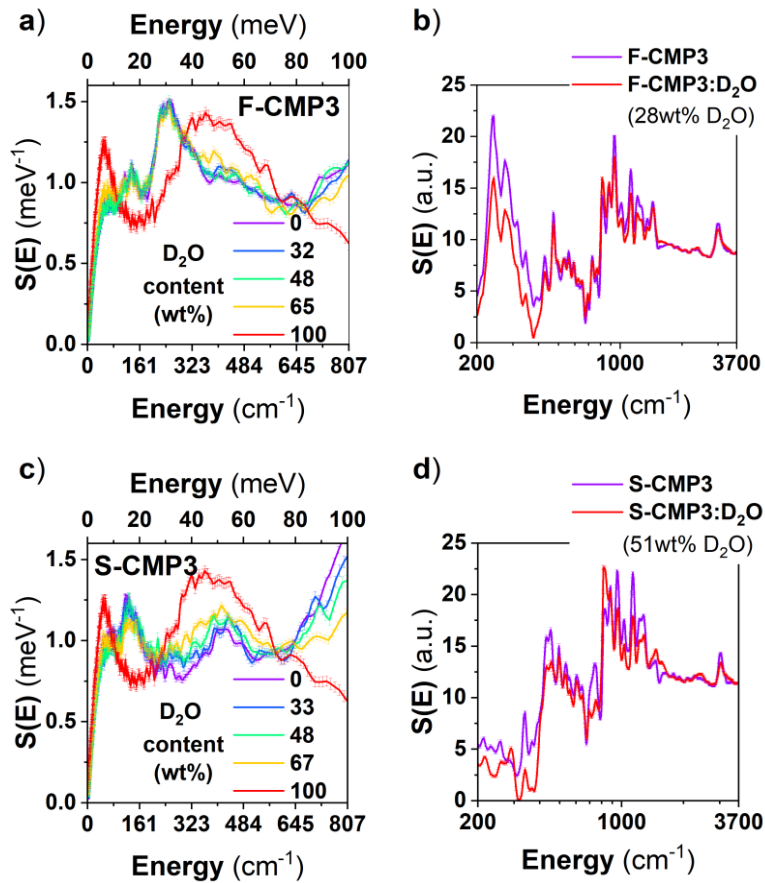

Figure 11. Room temperature GDOS of F-CMP3 (a), S-CMP3 (c), $D_2O$ and their mixtures, collected using the cold-neutron spectrometer IN6. The GDOS at 10K of F-CMP3 (b), S-CMP3 (d) and their mixtures with $D_2O$, using the hot-neutron spectrometer IN1-Lagrange, allowing to cover the full molecular vibrational range.

At low concentration (16 wt% for S-CMP3 and, 22 and 32 wt% for F-CMP3), the fits are not very satisfactory. Considering the large amount of bound water and the lower contrast between CMP and $H_2O$, the contribution from $H_2O$ to the QENS signal is significantly reduced and thus, it is impossible to perform a reliable fit. The amount of bound water is likely to be lower at lower concentration due to a reduced pressure. Furthermore, the CMP materials may exhibit an extra QENS contribution triggered by the hydration. In this context, in order to gain deeper insights, we performed further measurements with $D_2O$ for a contrast variation purpose between the CMP and water (Table S1 in Supporting Information). Interestingly, our attempt to model the low $H_2O$ concentration CMPs:$H_2O$ samples, and the $D_2O$-containing CMPs, with a weighted average of the dried CMP signal and the water signal extracted from the previous fit was successful (Supporting Information: Figures S2, S4 and S6). We find an amount of bound water close to the ones we obtained for $H_2O$ (Table 1); noting that for $D_2O$, the errors are too large to give meaningful numbers. Although the fits are reasonable for all the concentrations, it is improved for S-CMP3 and F-CMP3 with $D_2O$ at the highest concentrations, as compared to $H_2O$, when Q = 1.7 Å$^{-1}$ is not included in the data set. Diffractograms (Figure S9 in Supporting Information) exhibit the additional Bragg peak for CMPs mixed with $D_2O$. This supports the fact that a significant contribution from water is still probed.

Figure 11 shows the evolution of the vibrational spectra of both dried and wetted F-CMP3 and S-CMP3, from IN6 and IN1-Lagrange measurements. This time we make use of the unique contrast variation potential offered by neutron, and $D_2O$ is used instead of $H_2O$ for the wetted CMPs. The signal from water is no more of a dominant nature as $D_2O$ has a much lower neutron incoherent cross section than $H_2O$. Therefore, we expect that the changes between the dried CMP spectra and the wetted CMP spectra reflect, in this case, both the changes in the water and CMP spectra. Presently the observed changes are small. We consider the resolved IN1-Lagrange spectra and by subtracting dried and wetted F-CMP3 from dried and wetted S-CMP3 (Figure 12 a), respectively, we could highlight these changes. It is found that the differences due to hydration occur mainly in the energy range up to 1000 cm$^{-1}$ with two vibrational bands around 250 cm$^{-1}$ and 750 cm$^{-1}$ being the most strongly affected. In order to gain an insight into the nature of the modes concerned by these changes, we went a step further and we simulated the inelastic neutron spectra of the fluorene unit with the -CH$_2$ and -SO$_2$ group by adopting a DFT-based single-molecule approach. Figure 12 (b-c) compares measured and calculated neutron vibrational spectra of the dried CMPs. Interestingly, the agreement is found to be good. The single-molecule approach reproduced rather well the measured spectra, hence, reflecting the dominant intramolecular or simply the pronounced molecular aspect of the interaction within the CMPs. Having validated the calculated spectra, these can be used to spot some specific modes that could be relevant to the dynamics of the mass transfer we are reporting on. We found that these modes are related to the coupling of specific modes of the -C(CH$_3$)$_2$ and -SO$_2$ groups with out-of-plane motions of the backbones for the 250 cm$^{-1}$ band and with some modes of the benzene groups forming the fluorene unit for the 750 cm$^{-1}$ band. Figure 12 (a) highlights the impacted regions by these changes, and Table S8 gathers the mode frequencies and associated assignments of the CH$_2$ and SO$_2$ groups from our DFT-based single-molecule calculations.

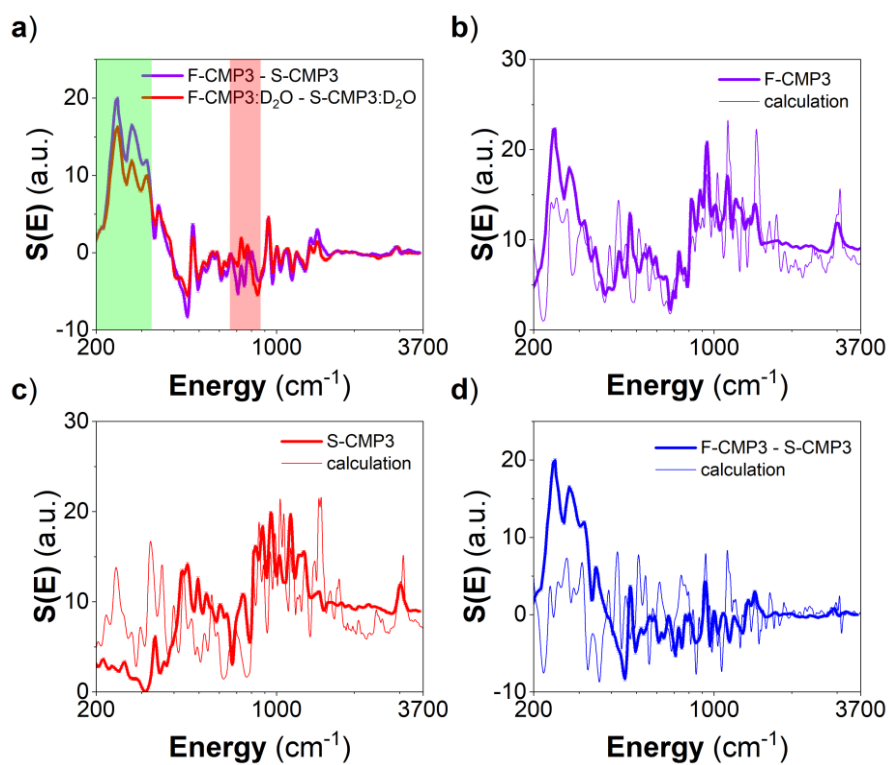

Figure 12. (a) Difference of measured GDOS of dried F-CMP3 and S-CMP3 compared with the difference of measured GDOS of D$_2$O-mixed F-CMP3 and S-CMP3. Comparison of measured and calculated GDOS of (b) dried F-CMP3, (c) dried S-CMP3, (d) the associated difference GDOS(F-CMP3)-GDOS(S-CMP3). The measured spectra were collected at 10K using the hot-neutron spectrometer IN1-Lagrange, allowing to cover the full molecular vibrational range. The calculated GDOS are DFT-based (0K), where a single molecule approach was adopted, neglecting any potential lattice effect (external degrees-of-freedom) and intermolecular interactions.

For hydrogen evolution applications, a hole scavenger is used in combination with water. We attempt to study the mass transfer of triethylamine (TEA) at 5 vol% in D$_2$O as used in previously reported hydrogen evolution measurements using QENS (Figure 13 a). Although no differences in the QENS spectra is observed for F-CMP3 with and without D$_2$O:TEA, the QENS spectra are different for S-CMP3 and S-CMP1 with and without D$_2$O:TEA. We can fit the D$_2$O:TEA with a similar model as for water (Supporting Information, Figure S7 and Table S7). The diffusion coefficient is lower than for bulk water and the residence time longer. Based on the fits of the QENS spectra with and without D$_2$O:TEA (Supporting Information, Figure S8), we do not observe within error bars of the measurement/fit (Figure 13 b), any differences between D$_2$O:TEA with and without the CMPs. Thus, we postulate that the TEA molecule does not enter the pores nor interact strongly with the CMPs but the difference in QENS spectra with and without D$_2$O:TEA is solely due to a superposition of the QENS signals of CMPs and D$_2$O:TEA. It It is worth noting that TEA is a rather large molecule and a smaller hole scavenger, with the appropriate energy level, might be beneficial to maximize the potential of those CMPs. This calls for further investigations to explore different hole scavengers.

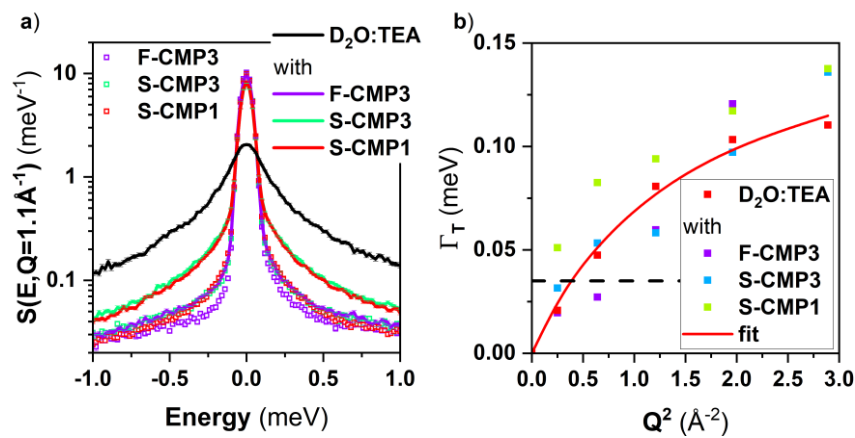

Figure 13. a) QENS spectra, at Q = 1.1 Å$^{-1}$, of F-CMP3, S-CMP3 and S-CMP1, the mixture of solvents D$_2$O:TEA and the CMPs mixed with D$_2$O:TEA. b) HWHM of the Lorentzian representing the translational diffusion of the mixture of solvents D$_2$O:TEA extracted from the fits as a function of Q$^2$, with a jump diffusion fit (solid line). The horizontal dashed line represents the instrumental resolution.

Conclusion

To summarize, the amount of bound water is significantly lower in S-CMP1 than in F-CMP3 and S-CMP3. This can be correlated with the difference in nodes between the two types of CMPs. The spiro link, which is likely to be stiffer, contributes to trapping water, resulting in an increase of the water described in this paper as "bound" water. The bound water and water exhibiting motions too slow to be captured by the spectrometer cannot be clearly differentiated here. The diffusion coefficient of water in S-CMP1 is nonetheless found to be significantly lower than free water indicating as in the case of S-CMP3 the presence of constrained water. All three CMPs present therefore a significant amount of trapped water but the range of dynamics of this trapped water differ. Indeed, water in F-CMP3 is either "bound" or free, while water in S-CMP1 is either constrained or free, S-CMP3 exhibits all the three behavior types of water.

The presence of the sulfone group, in S-CMP1 and S-CMP3, induces a change in translational motions of water accompanied by changes in the librational motions of water as observed by INS, while for F-CMP3, both translational and librational motions resemble the motion of bulk water. Therefore, it seems reasonable to infer that the stiffer spiro link contributes to a water trapping mechanism within the pores of the materials while the addition of the sulfone group leads to better interactions between the CMP surface and water and thus, induces slower dynamics of water.

We previously reported a much higher activity for S-CMP3 than for S-CMP1 and F-CMP3 that presented similar activities and all three CMPs were presenting larger activities than their linear analogue, pointing towards a benefit of porous materials.[45] F-CMP3, as measured by water sorption (Figure 1), absorb the least amount of water but similar activities as S-CMP1. The adsorbed water in F-CMP3 is mainly "bound" while is mainly constrained in S-CMP1. Thus, "bound" water seems the most beneficial water type for photocatalytic application. S-CMP3 benefits from both a large amount of "bound" water and a larger adsorption than F-CMP3 Photocatalytic activity is not only impacted by the mass transfer but also by the optoelectronic character of the materials. The nodes as well as the sulfone groups impact the electronic properties of the materials, and S-CMP3 was reported to have the lowest optical gap for instance.[45] The mass transfer has also to be balanced against the speed of the photocatalytic reaction. Therefore, it would not be plausible to draw a strong correlation between mass transfer and photocatalytic activity.[45] The addition of the hole scavenger TEA does not seem, presently, to lead to a strong interaction with the CMPs for the considered concentration. Further study with different concentrations and hole scavengers of different size are therefore needed.

## ASSOCIATED CONTENT

**Supporting Information**. Neutron incoherent cross sections of the samples, QENS spectra and associated fits with supplementary fitting parameters, neutron diffractograms, GDOS of CMPs mixed with D$_2$O and with D$_2$O:TEA obtained from IN6 measurements of all the measured samples. This material is available free of charge via the Internet at http://pubs.acs.org. The fit procedure detailed in the manuscript and in the SI was coded in a python script, which could be obtained from the corresponding authors upon a reasonable request.

## AUTHOR INFORMATION

Corresponding Author

*a.guilbert09@imperial.ac.uk
*zbiri@ill.fr

Author Contributions

A.A.Y.G. and M.Z. conceived and developed the project, wrote the neutron beamtime proposals, planned and performed the neutron experiments, treated and analyzed the neutron data. Y.B. and C.M.A. synthesized the materials and performed together with R.S.S. materials characterization. A.A.Y.G. and M.Z. wrote the manuscript with contribution from authors.

## ACKNOWLEDGMENT

The Institut Laue-Langevin (ILL) facility (Grenoble, France) is acknowledged for providing beam time on the IN6 and IN1-Lagrange spectrometers. A.A.Y.G. acknowledges the Engineering and Physical Sciences Research Council (EPSRC) for the award of an EPSRC Postdoctoral Fellowship (EP/P00928X/1). Y.B., C.M.A., and R.S.S from the group of Prof. Andrew I. Cooper thank EPSRC for funding (EP/N004884/1). Y.B. thanks the China Scholarship Council for a Ph.D. studentship.

ABBREVIATIONS

CMP – Conjugated Microporous Polymer
COF – Covalent Organic Framework
TEA – triethylamine
QENS – Quasi-Elastic Neutron Scattering
INS – Inelastic Neutron Scattering
GDOS – Generalized Density of States
HWHM – Half Width at Half Maximum